# Half-filled metal and molecular-orbital-mediated pairing in cuprate

Sixuan Chen[1†], Zhiheng Yao[1†], Ning Xia[1†], Shusen Ye[1†], Hongrui Zhang[1], Jianfa Zhao[2], Qingqing Liu[2], Changqing Jin[2], Shuo Yang[1,3,4], and Yayu Wang[1,3,4*]

[1]State Key Laboratory of Low Dimensional Quantum Physics, Department of Physics, Tsinghua University, Beijing 100084, China

[2]Beijing National Laboratory for Condensed Matter Physics, Institute of Physics, Chinese Academy of Sciences, Beijing 100190, China

[3]New Cornerstone Science Laboratory, Frontier Science Center for Quantum Information, Beijing 100084, China

[4]Hefei National Laboratory, Hefei 230088, China

[†]These authors contributed equally to this work.

[*]Corresponding author. Email: yayuwang@tsinghua.edu.cn

**Abstract:**

The cuprate high temperature superconductors exhibit anomalous momentum-space structures with antinodal gap and nodal arc in the underdoped regime, which evolves into a complete hole-type Fermi surface with a large Luttinger volume in the overdoped regime. The real-space electronic structure is also quite complex, as characterized by microscopic inhomogeneities and intertwined density wave orders. A valid theoretical framework for cuprates must explain these distinctive features across both reciprocal spaces in a unified manner. Here we show that doped holes in cuprate form localized electronic molecules consisting of $4a_0 \times 4a_0$ plaquettes, and each plaquette contains approximately two holes. The effective local doping level is thus around $p \sim 1/8$, which is sufficient to destroy the underlying antiferromagnetic order and more importantly, recovers the half-filled metallic state of the original $CuO_2$ plane. The restored Fermi surface, hosting one hole per unit cell, is consistent with experimental measurements and satisfies the Luttinger theorem. We then construct the momentum-space structure of the half-filled metal by taking into account the real-space configuration of electronic molecules. We show that the electronic potential with $4a_0$ periodicity imposed by the plaquettes and the quantum size effect of electronic molecules obliterate the nested antinodal Fermi surface sheets, leaving behind short arcs with coherent quasiparticles around the node. We propose that two doped holes in each plaquette occupy the shared molecular orbital and form a spin singlet, which can mediate the pairing of itinerant holes on the remnant Fermi surface of the half-filled metal. The electric dipole moment between the molecular orbitals and the dopant ions may also provide a novel attractive interaction between itinerant holes. This phenomenological model for pair formation between itinerant holes on the half-filled Fermi surface mediated by localized molecular orbitals resolves several core issues concerning the mechanism of superconductivity in cuprates.

**Introduction:**

High temperature superconductivity in cuprates originates from doping holes into the parent compound[1–3], which is a half-filled antiferromagnetic (AF) insulator[4] due to strong onsite Coulomb repulsion and superexchange coupling. Despite decades of extensive studies, there are still many puzzles in cuprates beyond current understanding. For example, angle-resolved photoemission spectroscopy (ARPES) on underdoped cuprates reveals the suppression of electron density of states (DOS) in the antinodal region of the Brillouin zone, generally known as a pseudogap[5–8]. With decreasing hole density, the pseudogap size becomes larger, and the remnant Fermi arc around the node shortens progressively[9–13]. In the optimally doped or overdoped regime, a complete hole-type Fermi surface (FS) takes into shape, as detected by ARPES[14–20], quantum oscillations[21–23], and quasiparticle interference (QPI) in scanning tunneling microscopy (STM) measurements[24–26]. Critically, the large FS pocket renders a Luttinger volume around $1 + p$, where $p$ is the density of doped holes, rather than simply $p$ expected for a doped Mott insulator. Therefore, even the fundamental Fermiology of cuprates has yet to be established convincingly, leaving the quest for the pairing mechanism without a valid starting point. The situation becomes even more complicated by the existence of intertwined charge and spin density wave orders that may compete or cooperate with superconductivity[27,28].

For a doped AF insulator, the real-space electronic structure may be more important than its momentum-space ($k$-space) counterpart because the physics is inherently local. We can gain valuable insights from the well-studied doped semiconductors, in which a charge dopant is modeled as a hydrogen-like atom[29]. The doped charge forms a bound state with the dopant ion by a screened Coulomb potential, allowing its energy level and spatial wavefunction to be derived by analogy to the hydrogen model. With increasing dopant density, the overlap of the wavefunction will broaden the impurity level into an impurity band, with numerous applications and new phenomena such as Anderson localization[30]. Similarly, a hole donor in cuprates can be regarded as a hydrogen-like atom embedded in the AF insulator background. Two neighboring dopants with overlapping wavefunction thus behave like a molecule with characteristic "molecular orbitals". As the doping level increases, the molecular orbitals will proliferate in space, pack into a dense network, and gradually lead to the superconducting state. STM is an ideal experimental probe for mapping out the electronic structure evolution in real space, which can also shed important new lights on the peculiar features of the reciprocal $k$-space.

**Localized molecular orbitals formed by doped holes:**

Our recent STM investigation explored the real-space electronic structure of alkali metal doped $Ca_2CuO_2Cl_2$ (CCOC) cuprate with dilute dopings[31]. Each alkali metal substitution for a Ca atom introduces an additional hole into the $CuO_2$ plane, which can be treated as a hydrogen atom in the AF insulator background. Figure 1 presents the experimental results on an insulating CCOC with alkali dopant density $p_A = 0.05$ (we use different symbols for doping due to the complications that will be discussed later). The topography in Fig. 1a appears inhomogeneous due to the random distribution of dopants, which can be identified by the bright tetramer encircling a Ca site. The low-energy $dI/dV$ map at $V_b = 50$ mV (Fig. 1b) reveals arrays of segmented plaquettes with internal short stripes, while the map at $V_b = 300$ mV (Fig. 1c) shows ladder rungs aligned perpendicular to the low-energy stripes. Both the shape and orthogonality of the stripe and ladder patterns are consistent with molecular orbitals developed by the linear combination of atomic orbitals of neighboring dopants with different relative phases[32]. We emphasize that the molecular orbitals here are different from those in real molecules because they originate from doped holes in the $CuO_2$ plane with strong electron correlation and AF exchange.

Figure 1d displays the d$I$/d$V$ curves obtained on an area with relatively extended electronic molecules, as marked by the rectangle in Fig. 1b-c. The gap exhibits a sharp V-shape with size around 10 meV, closely resembling the typical superconducting gap observed in Na-doped CCOC[33,34]. This indicates that Cooper pairing emerges first on the bright molecular orbitals, even in the insulating regime, as has been observed previously in $Bi_2Sr_2CuO_{6+\delta}$ (Bi-2201)[35]. Moreover, there is a positive correlation between the molecular orbital features and the superconducting properties[36,37,35,31]. The V-shaped gap is deeper and exhibits sharper coherence peaks on the bright molecular orbitals than on the dark space in between.

Figure 1e shows the linecut along the Cu-O bond direction in the Fourier transform (FT) of the map in Fig. 1b, which is consistent with previous reports on the checkerboard order[38–40]. There is a well-defined peak near 1/4 in the unit of $2\pi/a_0$, corresponding to the average size of $4a_0 \times 4a_0$ for each plaquette. The broad peak around 3/4 reflects the distance between neighboring stripes within the plaquette, which has a wide distribution around 1.3 $a_0$, or 4/3 $a_0$. An important finding is that each $4a_0 \times 4a_0$ plaquette contains approximately two holes, which can be estimated by dividing the number of doped holes by the number of plaquettes in the d$I$/d$V$ map. This conclusion holds for both underdoped CCOC[31] and Bi-2201 cuprates[35], two single-layer compounds that have been systematically investigated by STM. Consequently, the effective doped hole density for a plaquette is $p_{eff} \sim 2/(4 \times 4) = 1/8$. The d$I$/d$V$ maps reveal that the sample is spontaneously phase separated into dark regions without hole doping ($p_{eff} \sim 0$) and bright islands covered by plaquettes with $p_{eff} \sim 1/8$. This special type of phase separation has profound implications for the electronic structure and mechanism of pair formation in cuprates.

**Half-filled metal recovered by doped holes:**

Figures 2a-b display the d$I$/d$V$ maps at $V_b = 10$ mV and 200 mV for a CCOC with alkali dopant density $p_A = 0.10$. The stripe- and ladder-shaped molecular orbitals are nearly the same as those in the $p_A = 0.05$ sample, but now the plaquettes cover a larger portion of the surface. Consequently, the islands with local pairing establish long-range phase coherence, and the system enters the superconducting phase with $T_c = 14$ K. The superconductivity also exhibits a positive correlation with the structure of molecular orbitals, as shown by the height difference map[36] $H(\mathbf{r}) =$ d$I$/d$V(\mathbf{r}, 10$ mV$) -$ d$I$/d$V(\mathbf{r}, 0$ mV$)$ in Fig. 2c. Over such extended areas, the Bogoliubov QPI phenomenon can be revealed by the Z map, defined as $Z(\mathbf{r}, E = eV) = \frac{dI/dV(\mathbf{r},+V)}{dI/dV(\mathbf{r},-V)}$, as shown in Fig. 2d. The corresponding FT map in Fig. 2e exhibits scattering interference wavevectors characteristic for cuprates. The underlying FS contour derived from the octet model[41] constitutes an arc around the nodal direction[42,43], as illustrated in Fig. 2f, which quantitatively agrees with that measured directly by ARPES on a Na-doped CCOC with $p_A = 0.12$[9,17]. The Luttinger volume estimated by the FS area is around 1.03 (averaged hole number per unit cell), which represents a dramatic deviation from the expected value $p = 0.10$ for a doped Mott insulator, but is very close to 1 for a metal with half-filled band.

Putting all these experimental facts together, we arrive at the following physical picture. For areas covered by plaquettes, the effective doped hole density $p_{eff} \sim 1/8$ is sufficient to destroy the underlying AF order of the $CuO_2$ plane[44–46], as indicated by neutron scattering experiments on single crystals with 1/8 bulk doping[47–50]. Because the doped holes are localized by dopant potential and form static molecular orbitals, the only source for itinerant holes that constitute the large FS is the single hole on each Cu site occupying the $3d_{x2-y2}$ orbital, which is localized in the parent AF insulator. The destruction of AF order liberates these localized $3d$ holes, leading to a metallic state with half-filled band and Luttinger volume close to 1. We dub this phase a "half-filled metal".

The liberation of singly occupied holes on the Cu sites is easy to understand if the parent compound is a Slater-type AF insulator[51], in which the insulating phase is induced by the folding of the Brillouin zone by a pre-existing AF order. As soon as the AF order is destroyed by doped holes, the insulating gap closes and the $CuO_2$ plane immediately restores the half-filled metal state. Even if the parent compound is a true Mott insulator, the formation of molecular orbitals by doped holes with $p_{eff} \sim 1/8$ may weaken the Mottness to an extent that it becomes more appropriate to treat the $CuO_2$ plane from a half-filled metal limit. Irrespective of the exact microscopic process, the Luttinger volume of the resultant FS unequivocally indicates a half-filled Brillouin zone, which puts strong constraints on the origin of itinerant holes.

We must emphasize that the localized doped holes and the liberated itinerant holes are not entirely independent, and they collectively constitute a quantum many-body system. It is the interplay between them that dismantle the AF order in the $CuO_2$ plane. Nevertheless, as a first-order approximation, these two components can be treated separately. An analysis of the parent band structure indicates that the doped holes first occupy the oxygen $2p_{x,y}$ orbitals[52,53], and our previous studies demonstrated that the molecular orbitals can be constructed by oxygen $2p$ wavefunctions[54]. Therefore, the molecular orbitals can be regarded as an extended version of the oxygen $2p$ orbitals, co-existing with the liberated Cu $3d$ holes of the half-filled metal. This two-component picture represents a reasonable simplification of the three-band model for cuprates[55–58], with the $p_x$ and $p_y$ orbitals absorbed into the molecular orbitals. In fact, the Coulomb attraction from dopant ions, which was ignored in most theories, may significantly lower the doped hole energy compared to the Cu $3d$ hole. The hybridization between them may be much weaker than expected to justify the reduction into a single-band model[59]. The two-component picture is also reminiscent of the composite fermion model for the fractional quantum Hall effect, which reduces a formidable quantum many-body problem into two independent entities: magnetic flux and dressed electron[60,61]. In both cases, the interaction between these two components drives the fascinating quantum dancing of correlated electrons.

**Antinodal gap and nodal Fermi arc constructed from real-space patterns:**

A major obstacle for the recognition of a half-filled metal in underdoped cuprates is the existence of antinodal pseudogap and nodal Fermi arc[9], as opposed to a complete FS circle. We demonstrate in this session that these peculiar $k$-space features originate from the unique real-space configurations of molecular orbitals. As shown in Fig. 1b for the $p_A = 0.05$ sample, the electronic molecules consisting of $4a_0 \times 4a_0$ plaquettes are confined to small clusters rather than covering the whole surface. Such spatial patterns will have profound influences on the $k$-space electronic structure of the underlying half-filled metal through two different mechanisms.

First, the $4a_0 \times 4a_0$ plaquettes impose a quasi-periodic electronic potential that alters the $k$-space structure. As shown by ARPES measurements on Na-doped CCOC[9], the antinodal segments of the underlying FS are almost parallel to each other with a nesting wavevector $\sim 1/4$, which precisely corresponds to the $4a_0$ periodicity of the checkerboard order[38–40]. Traditional understanding attributes such electronic order to the charge density wave state, in which the FS nesting and electron-phonon coupling induce a permanent lattice distortion that gaps out the parallel FS sheets.

But here the physical process is the other way around. Our previous report demonstrated that even at very dilute doping $p_A = 0.03$ (Ref. 31), when the $k$-space structure is not well-defined, rod-shaped electronic molecules containing two or three $4a_0$ plaquettes already exist. The $4a_0$-periodic potential imposed by such pre-existing charge order can gap out the antinodal FS sheets with 1/4 nesting wavevector, leading to the pseudogap and remnant Fermi arc in the half-filled metal. It has a purely electronic origin without the involvement of lattice, which explains the absence of lattice distortion in CCOC and Bi-based cuprates exhibiting pronounced checkerboard order.

Second, in contrast to most theoretical approaches that assume an infinitely large periodic lattice for deriving the $k$-space structure, we observe that the rod-shaped electronic molecules are restricted to finite dimensions. Consequently, the underlying half-filled metal is subject to the quantum size effect, which may suppress certain states near the Fermi level and partially annihilate the FS. Because both the orientation of molecular orbitals and the direction of quantum confinement are along the Cu-O bond direction, the electronic states along the $(0, \pi)$ and $(\pi, 0)$ antinodal directions in $k$-space are expected to be strongly affected by the general principle of reciprocity.

To demonstrate the influence of molecular orbital configurations on the $k$-space structure, we perform numerical simulations based on a two-dimensional spinless tight-binding Hamiltonian on a square lattice with 100×100 sites and periodic boundary conditions. We start from a simple periodic lattice representing the original $CuO_2$ plane (Fig. 3a), and calculate the FS at half filling. The hopping integrals are set to: $t_1 = 1$ for nearest neighbors, $t_2 = -0.25$ for next nearest neighbors, and $t_3 = 0.2$ for second nearest neighbors. The calculated FS contour is shown in Fig. 3b, where the antinodal sheets are nearly parallel to each other with a nesting wavevector ~1/4, and the Luttinger volume is ~1. We then impose a periodic potential with $4a_0$ periodicity (Fig. 3c) to simulate the influence of plaquettes on the half-filled FS. The dark and light blue regions within each $4a_0$ plaquette have assigned chemical potentials $\mu = -0.32$ and $+0.25$, as indicated in the inset. Under such electronic potentials, the electron DOS in the antinodal region is significantly suppressed. The resulting FS (Fig. 3d) qualitatively resembles the ARPES data on CCOC with $p = 0.12$ (Ref. 9), characterized by a blurred antinodal region and a long Fermi arc around the node. To simulate the quantum size effect in the severely underdoped regime, we remove around 60% of the $4a_0$ plaquettes (Fig. 3e) by assigning a large local chemical potential $\mu = 1000$ to the black voids, and the remaining plaquettes form small disconnected clusters. The FS contour shown in Fig. 3f reveals a much stronger depletion of antinodal DOS and a reduced Fermi arc length with substantial linewidth broadening due to the quantum size effect, which is in excellent agreement with that measured by ARPES on CCOC with a similar doping level[9].

These simulations clearly demonstrate the influence of real-space patterns on the $k$-space electronic structure of cuprates that was overlooked in previous theoretical studies. The agreement between the single-particle simulations and experimental results implies that electron correlation in the half-filled metal is not too strong, and the ground state can still be described as a Fermi liquid. The quasiparticles near the FS are quite coherent as soon as the half-filled metal is recovered, as shown by the ARPES spectra on nodal Fermi arc[9]. It is the quasi-periodic potential of molecular orbitals that suppresses the antinodal DOS and makes it appear to be incoherent. The truncated FS also suggests that the Hall coefficient is no longer a valid probe for hole density, though in general the number of itinerant holes that contribute to transport increases with doping.

The picture that the mysterious pseudogap is mainly due to the impact of real-space electronic configurations on the $k$-space structure can explain most of the pseudogap phenomena that are related to electron DOS, such as the transport and thermodynamic properties. At elevated temperatures, the thermal activation energy helps the doped holes escape from the weak binding by dopant ions. With the melting of molecular orbitals, the quasi-periodic $4a_0$ potentials gradually vanish, so does the antinodal pseudogap. In the overdoped regime, the molecular orbitals smoothly transform into more uniform spatial features, as directly visualized by d$I$/d$V$ maps[26,37], leading to the closing of pseudogap. The densely doped holes in overdoped samples may form an impurity band and merge into the half-filled metal band, resulting in a Luttinger volume $1 + p$.

**Cooper pairing mediated by molecular orbitals:**

Now we have established a new framework about the basic electronic structure of cuprates: a network of localized molecular orbitals immersed in a half-filled metal on the $CuO_2$ plane. The pairing mechanism problem becomes well-defined, i.e., to find the attractive interaction between itinerant holes that opens a $d$-wave superconducting gap on the large half-filled metal FS[62]. A full FS circle is not necessary because the pairing can occur on two opposite Fermi arcs across the center of the Brillouin zone. Based on the observation that superconductivity is more pronounced on the static bright stripes, we propose that the pairing interaction is provided by the localized molecular orbitals. This picture is precisely the opposite to the resonating valence bond (RVB) theory for superconductivity in cuprates[1], which stated that the doped holes contribute itinerant carriers and the localized Cu spins provide the attractive interaction.

Based on our analyses, a microscopic theory for superconductivity in cuprates should include three terms in the Hamiltonian. The first term describes the formation of molecular orbitals by doped holes, mainly residing on the oxygen $2p$ orbitals and weakly bound by dopant ions. The second term describes the formation of a half-filled metal, mainly from the singly occupied holes on the Cu $3d$ orbital. The third term describes the interaction between them that leads to Cooper pairing on the large FS with $d$-wave symmetry. The exact microscopic origin of pair formation is beyond the scope of this paper; thus, we proceed in a purely phenomenological manner.

Figure 4 provides a series of cartoons to illustrate the formation of molecular orbitals and the possible pairing mechanism. Figure 4a depicts that when two dopants are in close proximity, the overlap of clover-shaped atomic orbitals leads to the formation of an electronic molecule in the shape of a single plaquette with size ~ $4a_0$. Two doped holes occupy the shared stripe-shaped molecular orbital and their spins form a singlet. In principle, both the charge and spin degrees of freedom of the molecular orbitals may provide the attractive interaction between two holes with opposite spins and momenta on the half-filled metal FS.

We start from the spin channel, which is generally believed to be the source of pairing in cuprate and other unconventional superconductors[63]. As illustrated in Fig. 4b, the spin singlet on the molecular orbital can mediate a Cooper pair through exchange coupling, analogous to the superexchange between two $3d$ electrons mediated by the $2p$ orbital of intermediary oxygen[64]. The main differences are that the two paired holes here are itinerant rather than localized, and the molecular orbital is much more extended than the $2p$ atomic orbital. This picture is consistent with the small Cooper pair size ~ 2 nm[65,66], which is comparable to the $4a_0$ size of a plaquette. The effective superexchange here should be smaller than $J$ ~ 130 meV between neighboring Cu spins in the parent compound due to the larger size of the plaquette, and its strength should determine the superconducting gap size, typically in the range of 10-50 meV. With increasing doping, the plaquettes aggregate into longer rod-shaped molecules and pack into clusters. The network of molecular orbitals may create entangled spin singlets, more like an RVB or quantum spin liquid state[67,68]. There are numerous theoretical models for Cooper pair formation mediated by RVB spin singlets or AF spin fluctuations[69–72,1,73,74]. For the simple situation described here with a square lattice and single hole band, most theories predicted that the $d_{x^2-y^2}$ pairing symmetry is favored.

The molecular orbitals also possess a unique charge density distribution, in which an extended hole cloud is loosely bound by dopant ions (with negative charges relative to the background). The resultant electric dipole moment is perpendicular to the $CuO_2$ plane, as shown schematically in Fig. 4c, which may also provide a pairing attraction. An itinerant hole near the electric dipole can polarize the hole cloud by Coulomb interaction, which in turn can attract another itinerant hole to its vicinity if the restoration of deformation is retarded compared to itinerant hole motion. For densely tiled molecular orbitals, the electric dipole moments assemble into an array of charged pendulums, which may develop their own bosonic mode that couples to itinerant holes and mediates the pairing. It sounds similar to the electron-phonon coupling in BCS superconductors, but the heavy lattice ions are not involved. Again, the attractive interaction is purely electronic in origin and can be much stronger than that mediated by phonons. This picture can also explain the small size of the Cooper pair, but it is unclear if it favors the *d*-wave pairing symmetry. The existence of electric dipole moment is quite rare in metal superconductors, so it could be another unique feature of cuprates that is responsible for the high $T_c$.

In either of these scenarios, the localized molecular orbitals behave like a network of pairing glues for the itinerant holes. In the underdoped regime, the closer packing of plaquettes with increasing doping leads to stronger phase coherence between superconducting islands, hence the increase of $T_c$ as described by the Uemura plot[75]. In strongly overdoped regime, the gradual melting of molecular orbitals not only closes the pseudogap, but also significantly weakens the pairing strength. The optimal doping represents a crossover when the spatial distribution of molecular orbitals is maximized over the whole sample. It is around $p = 0.15$ because at a global doping of $p = 1/8$, some regions remain uncovered by $4a_0$ plaquettes due to inhomogeneity[39,33,35], so a slightly higher doping level is needed to create the optimal environment for pairing.

**Discussions and conclusions:**

The results and arguments above provide a unified picture for understanding the Fermiology and pairing mechanism in cuprates at the phenomenological level. This framework yields several key points that are distinctively different from, if not totally opposite to, that expected for a doped Mott insulator. Central to this new paradigm is the pivotal role of molecular orbitals formed by the doped holes, as summarized below.

First, the wavefunction overlap between doped holes leads to the formation of molecular orbitals loosely bound by the dopant ions. There are approximately two holes in each $4a_0$ plaquette, so the effective local hole density $p_{eff} \sim 1/8$ is high enough to destroy the local AF order and recover the metallicity of the half-filled Cu sites. The density of liberated holes, one per unit cell, is consistent with the large Luttinger volume of the underlying FS.

Second, due to the quasi-periodic $4a_0$ potential provided by the molecular orbitals and the parallel FS sheets with 1/4 nesting wavevector, the antinodal states are strongly suppressed. The quantum size effect in the underdoped regime enhances such influence, and the nodal Fermi arc shrinks further. The pseudogap originates from the peculiar spatial configurations of molecular orbitals, and is not due to competing orders or precursor pairing.

Third, the two doped holes in each $4a_0$ plaquette occupy the shared molecular orbital and form a local spin singlet, which can provide an attractive interaction between two holes with opposite spin and $k$ on the FS of a half-filled metal, potentially via superexchange- or RVB-type mechanism. The electric dipole moment between the molecular orbitals and dopant ions may provide another novel type of pairing interaction through dynamic dipolar interaction with the itinerant holes.

The most counterintuitive part of this picture is that the doped holes do not contribute to itinerant carriers but instead act as local pairing glue. Most previous models reached an opposite conclusion because the dopant potentials, hence the localization tendency of doped holes, were largely overlooked. Importantly, STM results demonstrated clearly that the doped hole states are localized in space and form molecular orbitals. Furthermore, ARPES results showed that the itinerant holes form a large underlying FS with Luttinger volume always close to 1. The two-component picture proposed here represents the most reasonable model that satisfies both experimental constraints.

The final critical question is why two doped holes prefer to form molecular orbitals with a characteristic length scale around $4a_0$ in the first place. Due to the random distribution of dopants, in principle the electronic molecules formed by neighboring dopants could have widely varied size and shape. The self-organization into such a regular structure indicates that it is the energetically most favorable state. From the $t$-$J$ model viewpoint[76], the delocalization of doped holes lowers their kinetic energy, yet this process dismantles the local AF order and costs superexchange energy. Therefore, there is an optimal size to minimize the total energy. However, the validity of the $t$-$J$ model becomes questionable when the AF order is destroyed by an effective doping level $p_{eff}$ ~ 1/8. Instead, the large amount of kinetic energy released by the recovery of a half-filled metal becomes a much stronger driving force. Especially, the facts that $4a_0$ is nearly commensurate with the lattice and matches the 1/4 nesting wavevector of antinodal FS suggest that the formation of molecular orbitals in real-space and the formation of band structure in $k$-space are strongly interlocked. They cooperate in a highly synergistic way to reach such a magic situation: the half-filled metal is recovered to reduce kinetic energy, then the antinodal FS is gapped out to save more energy, and eventually the remnant Fermi arc is gapped out by superconductivity. The system takes advantage of nearly all possible mechanisms in condensed matter to reach the lowest energy state. This is a manifestation of the sophisticated quantum many-body physics in cuprates and the validity of the two-component picture. The intricate interplay between the localized molecular orbitals and the itinerant holes with truncated FS may also explain other exotic phenomena, such as the strange metal phase above $T_c$[77,78].

Back to the beginning, the phase separation of underdoped cuprate into AF insulator region ($p_{eff}$ = 0) and electronic molecule region ($p_{eff}$ ~ 1/8) not only achieves the lowest energy, but also substantially simplifies the problem. It circumvents the challenges of elucidating the complex phase diagram of a doped AF insulator with arbitrary hole density. Instead, we only need to focus on a simple situation with two doped holes in a $4a_0 \times 4a_0$ plaquette localized by the dopant ions. Remarkably, nothing competes with superconductivity in the half-filled metal after the AF order is destroyed by the doped holes. Phenomena such as the real-space checkerboard order and $k$-space antinodal pseudogap, previously viewed as competing factors, are shown here to be essential components behind the unique electronic structures required for the strong pairing. This is why superconductivity in cuprates is such a robust phase of matter. The overall story of high temperature superconductivity in cuprates vividly illustrates the power of emergence in a correlated electron system, both in real-space and $k$-space.

**Data Availability:** All raw and derived data used to support the findings of this work are available from the authors on request.

**Acknowledgement:** We thank X.T. Li, Y.Y. Peng, Z.Y. Weng, Q.K. Xue and B.F. Zhu for helpful discussions. Yayu Wang is supported by the Basic Science Center Project of NSFC (No. 52388201), the Innovation Program for Quantum Science and Technology (grant No. 2021ZD0302502), and the New Cornerstone Science Foundation through the New Cornerstone Investigator Program and the XPLORER PRIZE. Shuo Yang is supported by NSFC (grants No. 12174214 and No. 12475022) and the Innovation Program for Quantum Science and Technology (grant No. 2021ZD0302100). The work at IOP was supported by NSFC (grant No. 12204515), the National Key Research and Development Program of China (grants No. 2022YFA1403804, 2023YFA1406001), and the Young Elite Scientists Sponsorship Program of CAST (grant No. 2022QNRC001).

**Author contributions:** Y.W. initiated and supervised this project. J.Z., Q.L., C.J. prepared the CCOC single crystal. S.C., Z.Y., S. Ye, H.Z. carried out the STM experiments and data analysis. N.X. and S. Yang carried out numerical simulations. Y.W. wrote the manuscript with inputs from all authors.

**Competing interests:** The authors declare no competing interests.

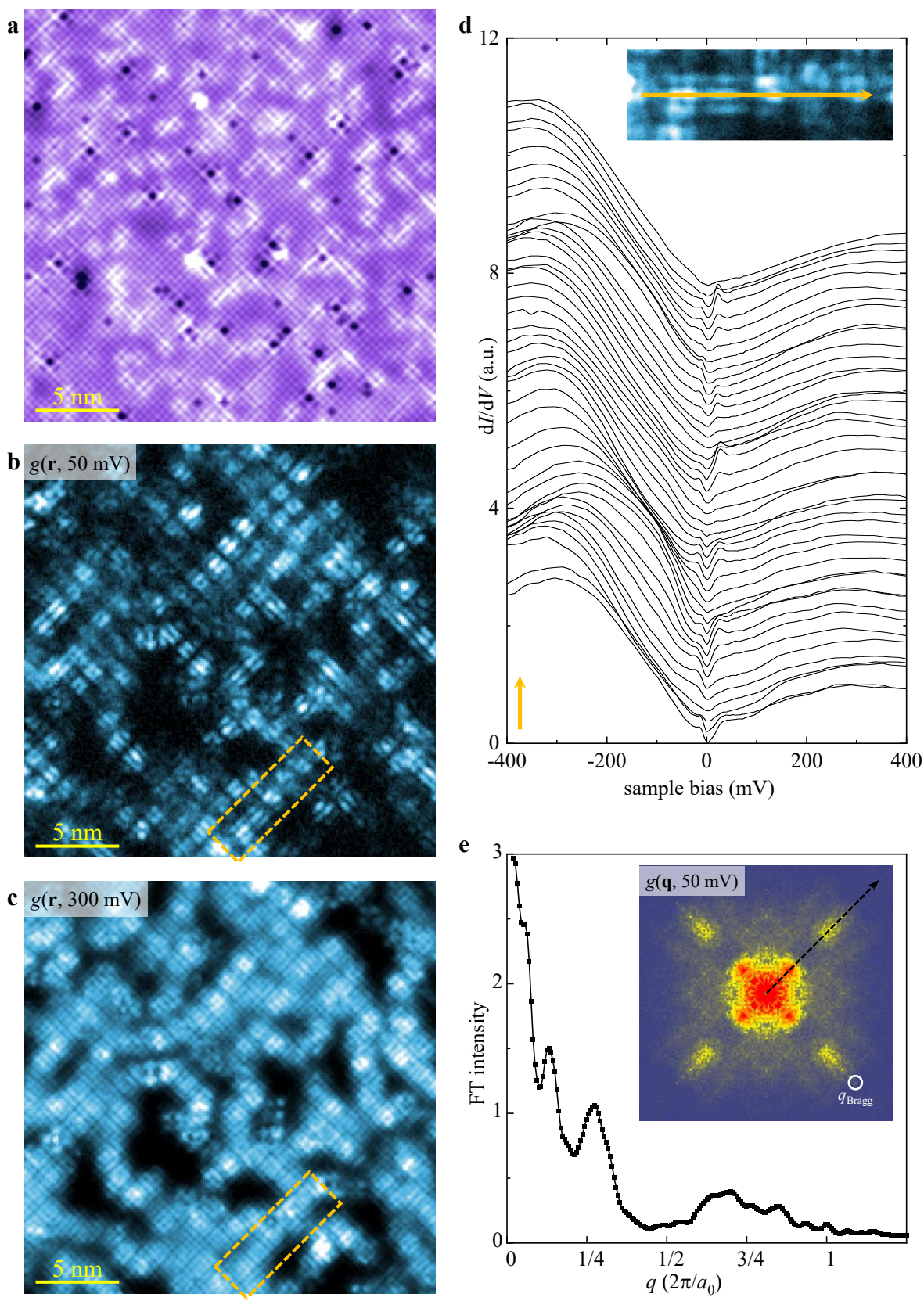

**Fig. 1| Topography and molecular orbitals in a $p_A$ = 0.05 CCOC. a**, Topographic image of a $p_A$ = 0.05 sample taken at $T$ = 5 K. **b-c**, d$I$/d$V$ maps at $V_b$ = 50 mV and 300 mV, respectively. Low-energy states in **b** present stripe arrays embedded in dark insulating background, whereas in **c** ladder patterns accompanied by sporadic clovers emerge. **d**, d$I$/d$V$ spectra along the line exhibit V-shaped superconducting gap, and the features are more pronounced on bright stripes. Inset: Zoomed-in d$I$/d$V$ maps on the dashed box area in **b** and **c**. **e**, Line profile of the FT map at 50 mV along the Cu-O bond direction, showing a sharp 1/4 $q_{Bragg}$ peak corresponding to the size of plaquettes. Inset: corresponding FT map with line profile direction marked by dashed arrow and the Bragg peak marked by white circle.

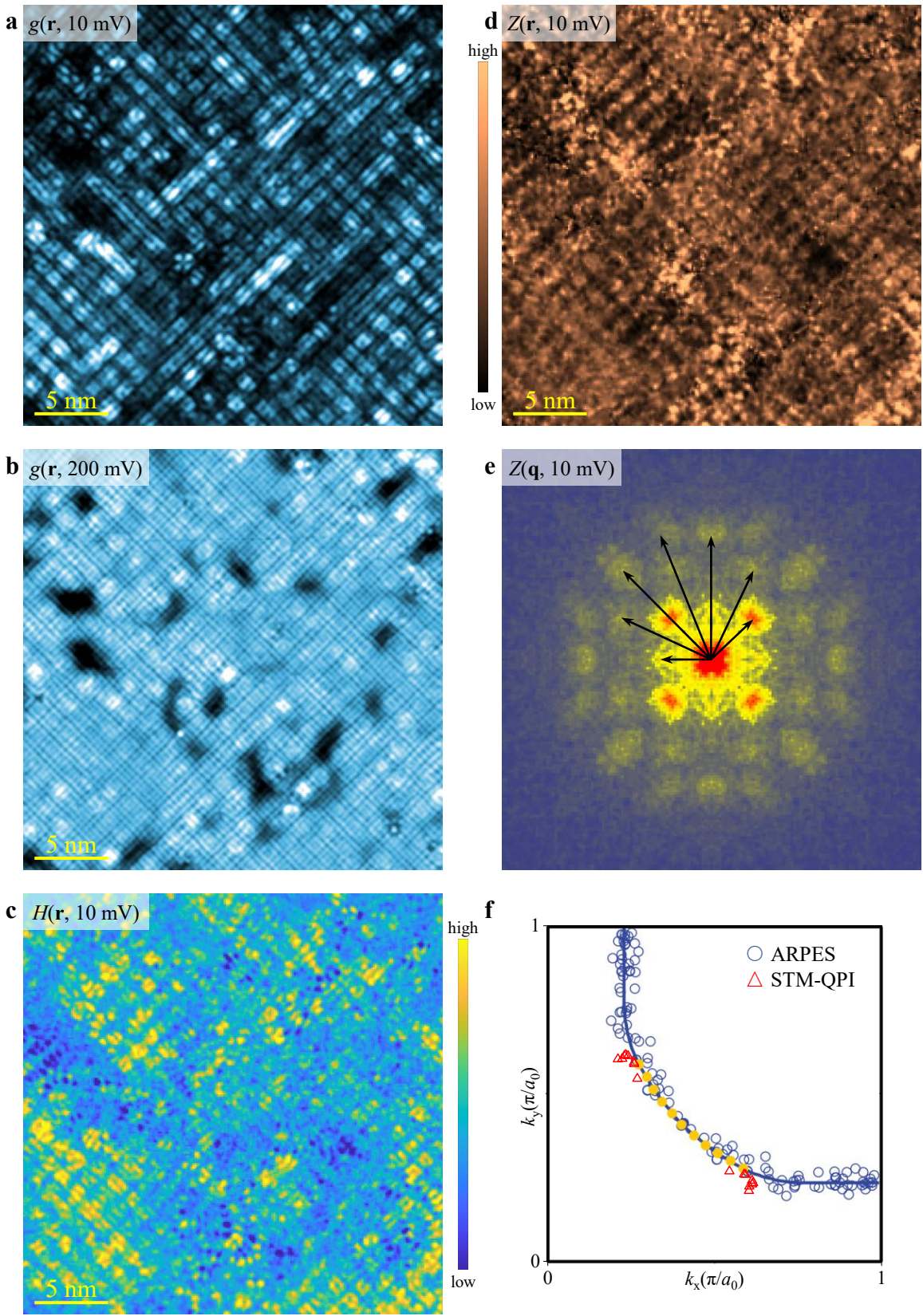

**Fig. 2| Real-space and *k*-space electronic structure in a $p_A = 0.10$ CCOC. a-b**, d$I$/d$V$ maps at $V_b$ = 10 mV and 200 mV, where closely tiled electronic molecules establish the checkboard order. **c**, The superconducting gap depth map $H(\mathbf{r}, 10\text{ mV})$, indicating the positive correlation between local strength of superconductivity and the molecular orbitals in **a**. **d**, The conductance ratio $Z(\mathbf{r}, 10\text{ meV})$ map exhibits wave-like quasiparticle interference patterns. **e**, Fourier Transform of **d**, revealing seven characteristic QPI wavevectors. **f**, Fermi surface derived from the octet model, in excellent agreement with the ARPES data adapted from Ref. 9, with a Luttinger volume close to 1.

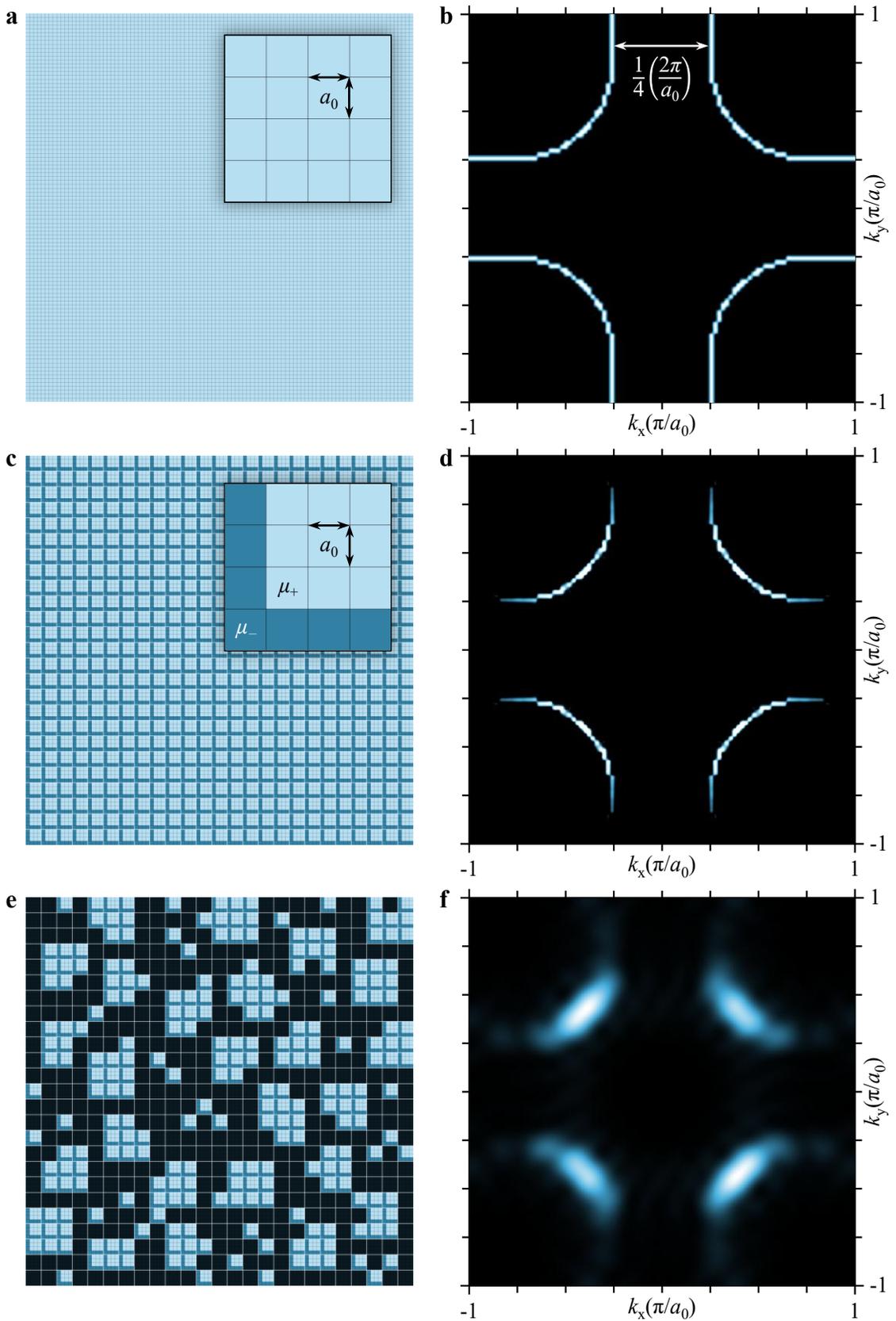

**Fig. 3| Numerical simulation of Fermi surface with varied real-space configurations. a**, Schematic of a 100×100 square lattice with lattice constant $a_0$. **b**, Fermi surface of a half-filled metal derived from the tight-binding model for **a**, with hopping integrals $t_1 = 1$, $t_2 = -0.25$, $t_3 = 0.2$. **c**, Schematic potential field with $4a_0 \times 4a_0$ periodicity. Inset: Zoom-in of a $4a_0$ unit showing local chemical potential configuration ($\mu_+ = 0.25$ and $\mu_- = -0.32$). **d**, Fermi surface derived for **c**, exhibiting DOS suppression at the antinodal regions. **e**, Potential field configuration with 60% plaquettes replaced by $\mu_b = 1000$ voids, simulating disconnected islands embedded in an insulating background, as observed in the $p_A = 0.05$ sample. **f**, Fermi surface derived for **e**, revealing short and broadened arcs around nodal direction.

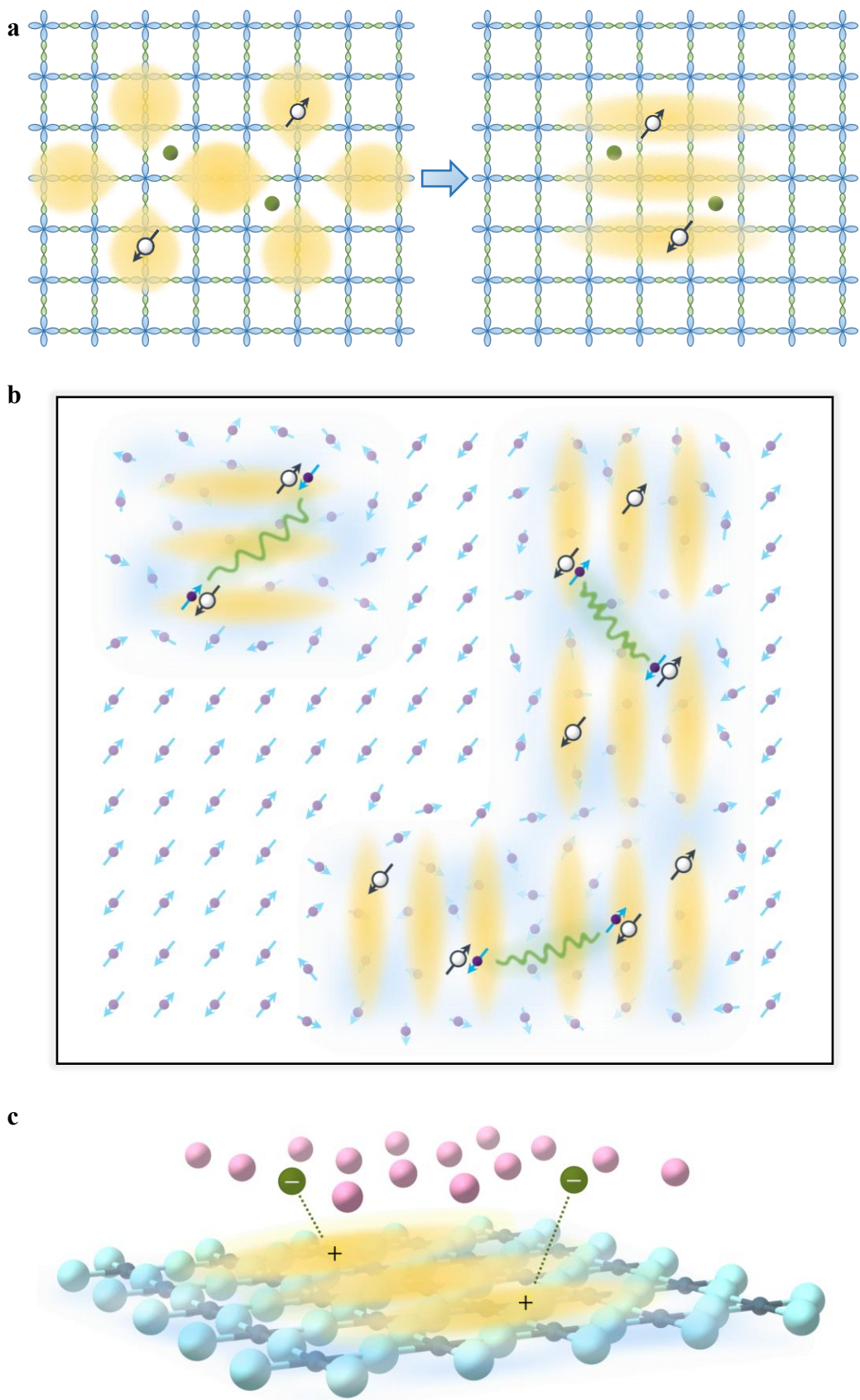

**Fig. 4| Formation of molecular orbitals and possible pairing mechanism. a**, Overlapping clover-shaped atomic orbitals of two neighboring dopants (olive dots) form molecular orbitals with a $4a_0$ plaquette and internal stripes. **b**, Schematic illustration of pair formation between two itinerant holes in the half-filled metal mediated by the spin singlets of molecular orbitals. **c**, Schematic formation of electric dipole moment between molecular orbital hole clouds and dopant ions, which may also mediate pairing of the itinerant holes.